\documentclass[aps,pra,reprint,showpacs,superscriptaddress,longbibliography]{revtex4-1} %
\usepackage{mathtools,amssymb,graphicx,units}
\usepackage[usenames,dvipsnames]{color}
\usepackage[plainpages=false,pdfpagelabels,colorlinks=true,linkcolor=red,urlcolor=blue,citecolor=blue,pdftitle={Title},pdfauthor={},pdfdisplaydoctitle=true,pdfduplex=DuplexFlipLongEdge]{hyperref}
\usepackage{verbatim}
\usepackage[english]{babel}

\usepackage[normalem]{ulem}
\usepackage{appendix}
\usepackage{bbm}
\newcommand{\beginsupplement}{%
        \setcounter{table}{0}
        \renewcommand{\thetable}{S\arabic{table}}%
        \setcounter{figure}{0}
        \renewcommand{\thefigure}{S\arabic{figure}}%
}
\graphicspath{{./fig/}}

\begin{document}
\title{Coexistence of ergodic and weakly ergodic states in finite-height Wannier-Stark ladders}
\author{Xingbo Wei}
\email{weixingbo@zstu.edu.cn}
\affiliation{Department of Physics and Key Laboratory of Optical Field Manipulation of Zhejiang Province, Zhejiang Sci-Tech University, Hangzhou 310018, China}
\author{Liangqing Wu}
\affiliation{Department of Physics and Key Laboratory of Optical Field Manipulation of Zhejiang Province, Zhejiang Sci-Tech University, Hangzhou 310018, China}
\author{Kewei Feng}
\affiliation{Department of Physics and Key Laboratory of Optical Field Manipulation of Zhejiang Province, Zhejiang Sci-Tech University, Hangzhou 310018, China}
\author{Tong Liu}
\email{t6tong@njupt.edu.cn}
\affiliation{Department of Applied Physics, School of Science, Nanjing University of Posts and Telecommunications, Nanjing 210003, China}
\author{Yunbo Zhang}
\email{ybzhang@zstu.edu.cn}
\affiliation{Department of Physics and Key Laboratory of Optical Field Manipulation of Zhejiang Province, Zhejiang Sci-Tech University, Hangzhou 310018, China}

\begin{abstract}
We investigate a single-particle in one-dimensional Wannier-Stark ladders with either a linear potential or a mosaic potential with spacing $\kappa=2$. In both cases, we exactly determine the critical energies separating the weakly ergodic states from ergodic states for a finite potential height. Especially in the latter case, we demonstrate a rich phase diagram with ergodic states, weakly ergodic states, and strongly Wannier-Stark localized states. Our results also exhibit that critical energies are highly dependent on the height of the ladder and ergodic states only survive at $E\approx0$ for the high ladder.
Importantly, we find that the number of ergodic states can be adjusted by changing the interval of the non-zero potential.
These interesting features will shed light on the study of disorder-free systems.
\end{abstract}

\maketitle

\section{Introduction}
Anderson localization~\cite{AndersonAbsence} is well-known as the phenomenon that the eigenfunction amplitude decays exponentially in space in disordered systems. It provides a foundational understanding of the insulating property of materials containing impurities. The scaling theory shows that Anderson localization is dimension-dependent~\cite{Abrahams,Thouless,EversAnderson,Mottthe}. In one- and two-dimensional systems without any symmetry, all states are Anderson localized states in the presence of arbitrarily weak uncorrelated disorder~\cite{Abrahams,Thouless}. While in three dimensions, there is a localization transition from extended states to Anderson localized states as the disorder strength increases. In this three-dimensional system, Anderson localized states and extended states coexist, and they are separated by critical energies, dubbed as mobility edges~\cite{EversAnderson}. Since then, various uncorrelated disordered systems and quasi-periodic (quasi-randomness) systems have been discovered to display the existence of mobility edges~\cite{Aubry,Harper,SarmaMobility,Wangone,LiuGeneralized,LiuMobility,LiMobility,XiaExact,Mondaini,Ganeshan,Miguel,lin2023general,edgestong,Dengprl,Biddle,Qi-Bo}, and part of them have been observed in the experiments~\cite{Semeghini,PasekAnderson,uschen,LinTopological}.

\begin{figure}[htbp]
	\includegraphics[width=0.9\columnwidth,height=0.7\columnwidth]{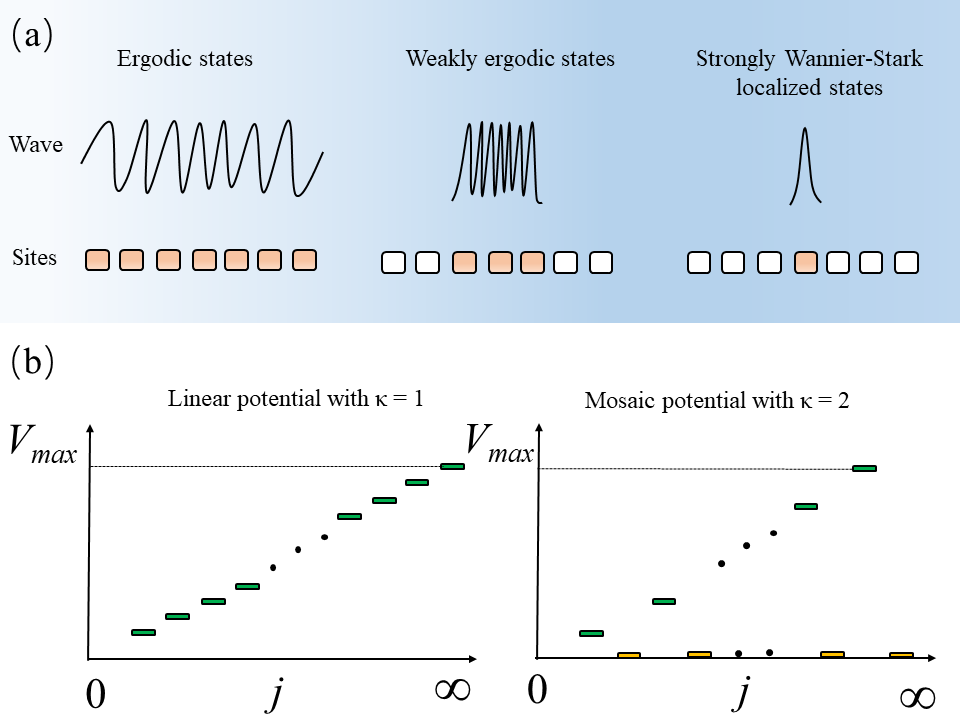}
	\vspace{-0.4cm}
	\caption{The schematic plot of wave functions and potentials. (a) Typical wave functions for ergodic states, weakly ergodic states, and strongly Wannier-Stark localized states. (b) The linear potential with spacing $\kappa=1$ and the mosaic potential with spacing $\kappa=2$. In (a), rectangles represent lattice sites and orange filled ones represent occupied sites by the wave function. In (b), $j$ is the site index and $V_{max}$ represents the maximum value of the potential. The green and yellow lines indicate that the potential energy is non-zero and zero, respectively. It is necessary to emphasize that the critical energies obtained in our work separate ergodic states and weakly ergodic states in the spectrum.
	}
	\label{fig0}
\end{figure}

Recently, it has been reported that mobility edges also exist in disorder-free systems with mosaic modulations for spacing $\kappa>1$~\cite{Dwiputrasingle}. However, Ref.~\cite{AbsenceLonghi} shows the opposite conclusion to Ref.~\cite{Dwiputrasingle}, which analytically proves that there is no mobility edge in the strict sense in the mosaic Wannier-Stark lattices and points out that the Avila's theory ~\cite{Avila} is not applicable to these systems. The corresponding experiment is also reported in Ref.~\cite{gaoobservation}.  To illustrate the motivation of this work, we refer to the definition of Wannier-Stark localization. For a model in the presence of a linear potential, the eigenstates are the well-known Wannier-Stark states $|\Psi_m\rangle=\sum_j \mathcal{J}_{j-m}(2t /V)|j\rangle$ ~\cite{Fukuyamatightly}, where $\mathcal{J}_{j-m}$ are the Bessel functions of the first kind, $j$ and $m$ represent the site index and energy level index, respectively. The properties of the Bessel functions show that $\mathcal{J}_{j-m}$ are mainly localized in the interval $|m-j|<2t/V$
~\cite{hartmann2004dynamics}. In the thermodynamic limit with the system size $L\rightarrow\infty$, $|m-j|<2t/V\ll L$ for arbitrarily finite $V$, thus all states are Wannier-Stark localized states~\cite{hartmann2004dynamics,gluck2002wannier,van2019bloch,zhuStatic}. This case corresponds to the infinite-height Wannier-Stark ladder due to the height of the ladder $VL\rightarrow\infty$. Different from it, when the Wannier-Stark ladder is finite-height, the extended states can also exist~\cite{Mendez,Voisin,Mendez2}. Thus the height of the ladder is important for Wannier-Stark localization. In this work, we mainly focus on the case of the finite-height Wannier-Stark ladders to identify different phases. To introduce the different phases observed in this work, we explicitly define that the ergodic state corresponds to the wave function distributing throughout the space, the weakly ergodic state corresponds to the wave function living on a finite fraction of all the sites with the fractal dimension $D=1$ ~\cite{Masudul}, and the strongly Wannier-Stark localized state corresponds to the wave function localizing in about a single site~\cite{qilocalization,YanxiaA,qilocalization,YanxiaA}, as shown in Fig.~\ref{fig0} (a).

\section{Model}
We investigate a quantum particle in a disorder-free chain with open boundary conditions, which is described by the following Schr\"{o}dinger equation~\cite{Berezine}
\begin{equation}\label{h}
t\left(\psi_{j+1}+\psi_{j-1}\right)+V_j\psi_{j}=E\psi_{j},
\end{equation}
where $\psi_{j}$ is the amplitude of the wavefunction at the site $j$. $E$ is the eigenvalue. We set the nearest-neighbor hopping strength $t\equiv1$ as the energy unit and $V_j$ is the site-dependent potential, which reads
\begin{equation}\label{pet}
V_j= \begin{cases}V j, & j=\kappa i \\ 0, & \text { otherwise }\end{cases},
\end{equation}
where $V$ is the strength of the linear potential. $\kappa$ adjusts the spacing of sites with the non-zero potential. Typically, we choose $\kappa=1$ and $\kappa=2$ in this work, corresponding to the linear potential and the mosaic potential~\cite{Wangone,Dwiputrasingle}, respectively. The schematic plots of potentials are shown in Fig.~\ref{fig0} (b). Without specification, $i=0,1,2, \ldots, (N-1)$ represents the location of the supercell. Each supercell includes $\kappa$ sites, thus the chain length is expressed as $L=\kappa N$. The maximum value of the potential is defined as $V_{max}=\kappa V(N-1)$. In the present work, we use finite-height ladders~\cite{Mendez,Voisin,Mendez2}, i.e. $V_{max}$ does not depend on the system size and $V\propto1/(N-1)$.

For $\kappa=1$, Eq. \eqref{h} is reduced to the Wannier-Stark model~\cite{Wannier,Fukuyama}, whose eigenstates are all Wannier-Stark localized states for arbitrarily finite $V$~\cite{hartmann2004dynamics,gluck2002wannier,van2019bloch,zhuStatic}. For $\kappa=2$, disorder-free mobility edges are reported recently~\cite{Dwiputrasingle}. Remarkably, these two models have already been realized in experiments in the superconducting qubit and nanophotonic device systems~\cite{guo2021observation,gaoobservation}.

\section{Lyapunov exponent and critical energies}
We start from Eq. \eqref{h} and transform it into the transfer matrix form as
\begin{equation}\label{matrix}
\left[\begin{array}{c}
\psi_{j+1} \\
\psi_j
\end{array}\right]=T_j\left[\begin{array}{c}
\psi_j \\
\psi_{j-1}
\end{array}\right],
\end{equation}
where $T_j$ is given by
\begin{equation}\label{T_j}
\begin{aligned}
T_j & =\left(\begin{array}{cc}
E-V_j & -1 \\
1 & 0
\end{array}\right).
\end{aligned}
\end{equation}
For convenience, we abbreviate Eq. \eqref{matrix} as $\Psi_{j}= T_j \Psi_{j-1}$. The transfer matrix of the supercell $\tilde{T}_i$ is composed of $\kappa$ small transfer matrices ${T}_j$, thus it can be written as
\begin{equation}\label{T_i}
\begin{aligned}
\tilde{T}_i &= \prod_{j=\kappa i}^{\kappa i+\kappa-1} \left(\begin{array}{cc}
E-V_j & -1 \\
1 & 0
\end{array}\right) \\
& =\left(\begin{array}{cc}
E-V\kappa i  & -1 \\
1 & 0
\end{array}\right)\left(\begin{array}{cc}
E & -1 \\
1 & 0
\end{array}\right)^{\kappa-1}.
\end{aligned}
\end{equation}
The Lyapunov exponent indicates the exponential rate of growth of the transfer matrix product, which is defined as
\begin{equation}\label{e5}
\begin{aligned}
\gamma(E) &=\lim _{L \rightarrow \infty} \frac{1}{L} \ln (||\prod_{j=0}^{L-1} T_j||) \\
&= \lim _{L \rightarrow \infty} \frac{1}{L} \ln (||\prod_{i=0}^{N-1} \tilde{T}_i||), \\
\end{aligned}
\end{equation}
where $||$ $\cdot$ $||$ is the norm of the matrix, determined by the maximum of the absolute value of eigenvalues. For localized states, $\gamma(E)>0$, whereas for non-localized states $\gamma(E)=0$~\cite{LiMobility,ZhangLyapunov}. The Lyapunov exponent is widely used in the studies of Anderson localization and mobility edges~\cite{LiMobility,ZhangLyapunov,WangDuality,CaiExact,JiangMobility,LiuExact,LiuYanxia}. In the present work, we use the Lyapunov exponent to determine the critical energies separating ergodic states and weakly ergodic states, since the weakly ergodic states in this work have $\gamma>0$, which is different from ergodic states with $\gamma=0$. Since the height of the Wannier-Stark ladder is finite in the present work, $\tilde{T}_i\approx \tilde{T}_{i+1}$ for $L\rightarrow\infty$ according to Eq. \eqref{T_i}, which results in that the transfer matrix product can be converted to $||\prod_{i=0}^{N-1} \tilde{T}_i||=||\phi_{N-1}\Lambda_{N-1}\phi_{N-1}^\dag\phi_{N-2}\Lambda_{N-2}\phi_{N-2}^\dag \ldots \phi_{0}\Lambda_{0}\phi_{0}^\dag ||\approx ||\phi_{N-1}\prod_{i=0}^{N-1} \Lambda_{i}\phi_{0}^\dag||\approx ||\prod_{i=0}^{N-1} \Lambda_{i}||=\prod_{i=0}^{N-1} || \tilde{T}_i||$, where $\phi$ and $\Lambda$ are matrices composed of the eigenstates and eigenvalues of the transfer matrix $\tilde{T}$, respectively. Thus we obtain that the Lyapunov exponent is approximated as
\begin{equation}\label{e6}
\begin{aligned}
\gamma(E) \approx \lim _{L \rightarrow \infty} \frac{1}{L} \ln (\prod_{i=0}^{N-1} ||\tilde{T}_i||). \\
\end{aligned}
\end{equation}
By a direct computation, $\gamma(E)$ for $\kappa=1$ and $\kappa=2$ can be obtained as
\begin{equation}\label{lv}
\begin{aligned}
\gamma(E)&\approx \lim _{L \rightarrow \infty} \frac{1}{L}\sum_{i=0}^{N-1}\ln (\text{max}\{ |\varepsilon_1|, |\varepsilon_2| \}) \\
&=\lim _{L \rightarrow \infty} \frac{1}{L}\sum_{i=0}^{N-1}\Bigg| \ln(|\frac{ -\mu_i +\sqrt{\mu_i^2-4}}{2}|) \Bigg|, \\
\end{aligned}
\end{equation}
where $\varepsilon_1$ and $\varepsilon_2$ are eigenvalues of $\tilde{T}_i$, $\varepsilon_1=1/\varepsilon_2$ due to the determinant $|\tilde{T}_i|=1$. $\mu_i=E-V i$ for $\kappa=1$ and $\mu_i=E^2-2EVi-2$ for $\kappa=2$. Evidently, $\gamma(E)=0$, the condition gives $|\frac{ -\mu_i +\sqrt{\mu_i^2-4}}{2}|=1$ for any $i$, which can be satisfied if $|\mu_i|\leq2$ for any $i$.
Here $|\mu_i| = 2$ corresponds to the critical energies and $|\mu_i|<2$  gives the energy interval for ergodic states.
Since the non-zero potential is linear, $|\mu_i|\leq2$ for any $i$ can be guaranteed as long as both the head ($i=0$) and end ($i=N-1$) of the ladder satisfy $|\mu_i|\leq2$. Thus, according to $|\mu_{0}(E_{c1})|=2$ and $|\mu_{N-1}(E_{c2})|=2$, we can obtain that critical energies for $\kappa=1$ are at
\begin{equation}\label{k1}
E/t=\left\{\begin{aligned}
&2, \\
&-2+V_{max}/t,\\
\end{aligned}\right.
\end{equation}
and those for $\kappa=2$ are at
\begin{equation}\label{k2}
E/t=\left\{\begin{aligned}
&0, \\
&2, \\
&\frac{V_{max}/t-\sqrt{(V_{max}/t)^2+16}}{2},\\
&V_{max}/t,\\
\end{aligned}\right.
\end{equation}
It needs to be emphasized that the above derivation is based on $\tilde{T}_i\approx \tilde{T}_{i+1}$, which requires the height of the Wannier-Stark ladder to be finite.

To verify the predictions of Eq. \eqref{k1} and Eq. \eqref{k2}, we numerically calculate the Lyapunov exponent by the following method from the original definition~\cite{ZhangLyapunov}
\begin{equation}\label{11}
\begin{aligned}
\gamma(E)&=\lim _{L \rightarrow \infty} \frac{1}{L} \ln (|\Psi_{L-1}|/ |\Psi_0|) \\
&= \lim _{L \rightarrow \infty} \frac{1}{L}\ln (\frac{|\Psi_{L-1}|}{|\Psi_{L-2}|} \frac{|\Psi_{L-2}|}{|\Psi_{L-3}|} \ldots \frac{|\Psi_1|}{|\Psi_0|}) \\
&= \lim _{L \rightarrow \infty} \frac{1}{L}\sum_{j=0}^{L-2} \ln (\frac{|\Psi_{j+1}|}{|\Psi_{j}|}),
\end{aligned}
\end{equation}
where $|\Psi_j|=\sqrt{|\psi_{j+1}|^2+|\psi_j|^2}$ is the norm of the vector. Note that the Lyapunov exponent here is over the entire chain. For weakly ergodic states, the wave function lives on a finite fraction of all the sites with a higher than exponential decay, and one may investigate the site-dependent Lyapunov exponent to characterize its properties as shown in Appendix E. In this work, the Lyapunov exponent of the entire chain works well to separate ergodic states from other states. In detail, we first choose a normalized vector
\begin{equation}
\begin{aligned}
\Psi_0 = \left(\begin{array}{c}
\sqrt{2}/2 \\
\sqrt{2}/2
\end{array}\right),
\end{aligned}
\end{equation}
as the initial vector and set the initial Lyapunov exponent to $\gamma(E)=0$. Then, we iterate $j$ from zero to $L-2$ to calculate the Lyapunov exponent according to Eq. \eqref{11} by the following steps:
\begin{itemize}
\item[1.] normalize the vector $\Psi_{j}$ at the site $j$ and the normalized vector is still denoted as $\Psi_{j}$;

\item[2.] calculate the next vector $\Psi_{j+1}= T_{j+1} \Psi_{j}$ by Eq. \eqref{matrix};

\item[3.] calculate the Lyapunov exponent $\gamma(E)= \gamma(E) + \frac{1}{L}\ln (|\Psi_{j+1}|/|\Psi_{j}|)= \gamma(E) + \frac{1}{L}\ln |\Psi_{j+1}|$, since $\Psi_{j}$ has been normalized in the first step.
\end{itemize}

Here we do not adopt the analytical method in Ref.~\cite{AbsenceLonghi}, since the weakly ergodic states separated from ergodic states by critical energies are localized at the head or the end of the lattice chain. This causes the results in Ref.~\cite{AbsenceLonghi} to be no longer available in the present work, details are shown in the Appendix.

\section{results}
\begin{figure}
	\includegraphics[width=1.0\columnwidth,height=0.6\columnwidth]{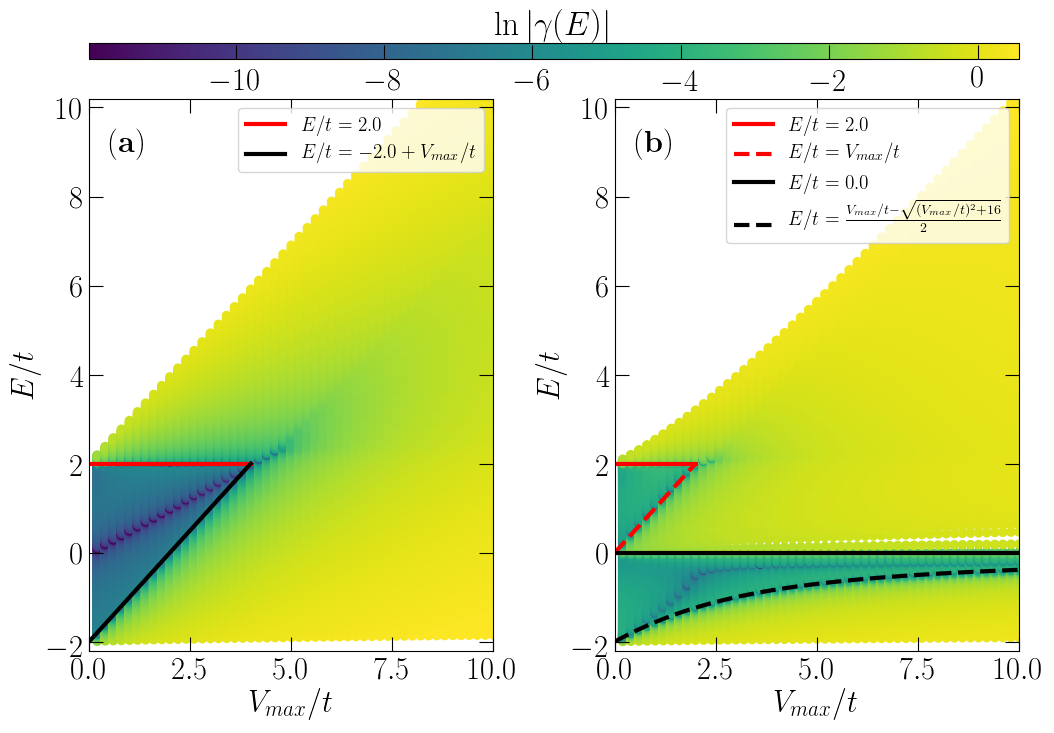}
	\vspace{-0.4cm}
	\caption{Spectra and critical energies as a function of $V_{max}$. Color represents the value of $\ln|\gamma(E)|$. The dark (light) color region corresponds to the ergodic (weakly ergodic) region. The value of $V_{max}$ here is far from meeting the requirements for achieving strong Wannier-Stark localization. The solid and dashed lines in red and black mark the critical energies separating ergodic states from weakly ergodic states. (a) $\kappa=1$ and (b) $\kappa=2$. $L=500$.
	}
	\label{fig1}
\end{figure}
The main results are shown in Fig.~\ref{fig1}, in which we use the original Lyapunov exponent defined by Eq. \eqref{11} to characterize critical energies. In Fig.~\ref{fig1} (a), we consider the linear potential for $\kappa=1$. Two critical energies are at $E/t=2$ and $E/t=-2+V_{max}/t$ for $V_{max}/t<4$, which agrees with Eq. \eqref{k1}.
As $V_{max}$ increases, the region of ergodic states is compressed gradually. For $V_{max}/t>4$, ergodic states disappear and all states in the spectrum are weakly ergodic states in Fig.~\ref{fig1} (a). Further enhancing the strength of the potential, the system enters the strong Wannier-Stark localization region, where the particle localizes in about a single lattice~\cite{Dwiputrasingle,qilocalization,YanxiaA}. In the thermodynamic limit, $V_{max}\rightarrow\infty$ for arbitrarily finite $V$, thus it can be concluded that arbitrarily finite $V$ can induce the system to enter the Wannier-Stark localization, which is consistent with Ref.~\cite{hartmann2004dynamics,gluck2002wannier,van2019bloch,zhuStatic}. It is worth mentioning that from the perspective of energy conservation, ergodic states and weakly ergodic states can also be understood as scattered states and bound states~\cite{Phillips}, thus critical energies can also be obtained by analyzing eigenvalues~\cite{zhuStatic}. In Fig.~\ref{fig1} (b), we consider the mosaic potential for $\kappa=2$. There are four critical energies for $V_{max}/t<2$, and two of them disappear for $V_{max}/t>2$, only $E/t=0$ and $E/t=(V_{max}/t-\sqrt{(V_{max}/t)^2+16})/2$ survive. These two critical energies approach each other as $V_{max}$ increases. Thus for a large $V_{max}$, ergodic states only exist at $E/t\approx0$.

\begin{figure}[htbp]
	\includegraphics[width=1.0\columnwidth,height=0.8\columnwidth]{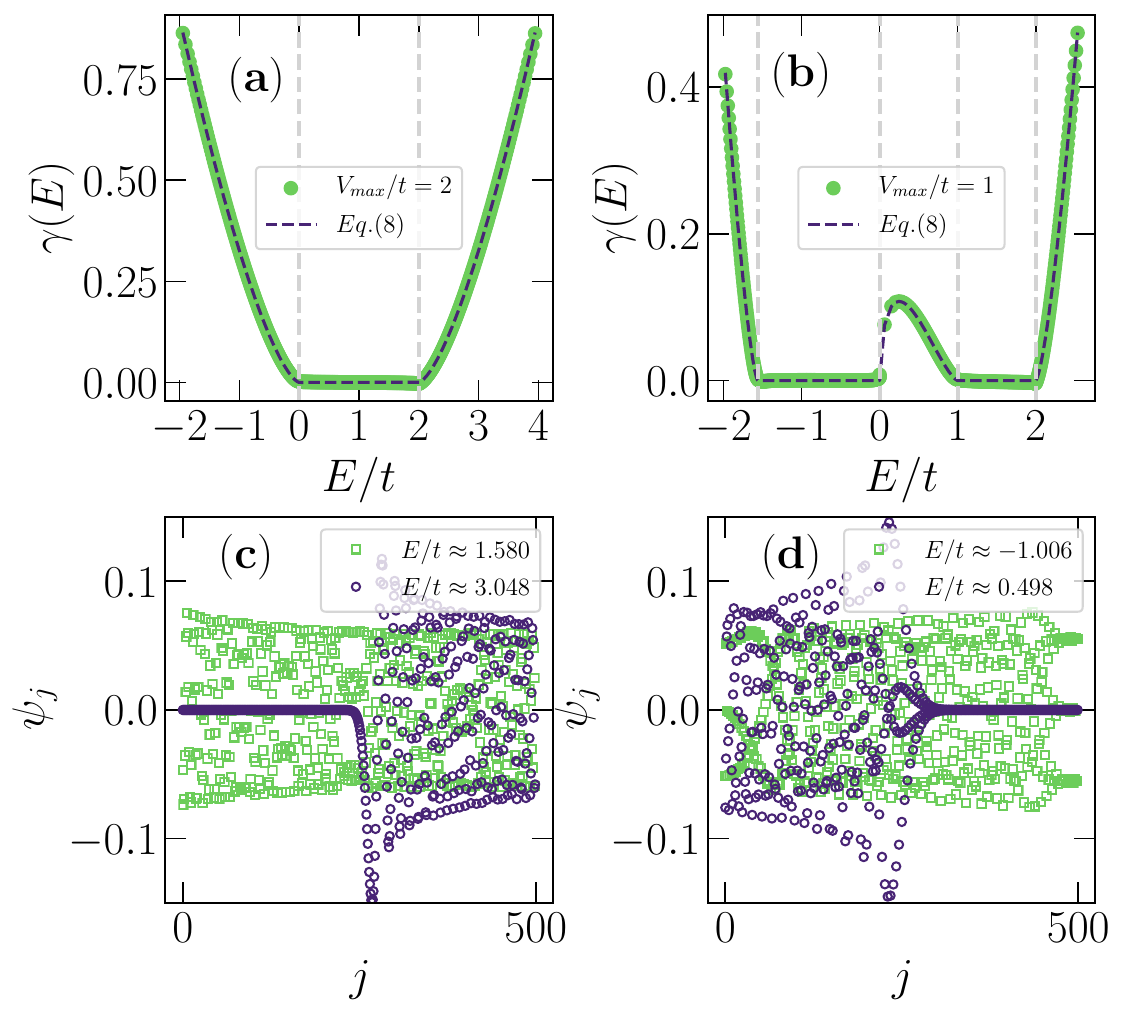}
	\vspace{-0.4cm}
	\caption{ Lyapunov exponents $\gamma(E)$ and wave functions $\psi_j$ for $\kappa=1$  (left panels) and $\kappa=2$  (right panels).  The grey dotted lines mark critical energies at $E/t=0$ and $E/t=2.0$ in (a), whereas those in (b) mark critical energies at $E/t=\frac{1-\sqrt{17}}{2}$, $E/t=0$, $E/t=1$ and $E/t=2.0$. $L=500$. $V_{max}/t=2.0$ in the left panels and $V_{max}/t=1.0$ in the right panels.
    }
	\label{fig2}
\end{figure}

In Fig.~\ref{fig2}, we show typical Lyapunov exponents and wave functions for $\kappa=1$ (left panels) and $\kappa=2$ (right panels). In both cases, the Lyapunov exponent $\gamma(E)=0$ for ergodic states, whereas $\gamma(E)>0$ for weakly ergodic states in Fig.~\ref{fig2} (a) and (b). The numerical solution of the Lyapunov exponent coincides well with Eq. \eqref{lv}, indicating that the approximation $\tilde{T}_i\approx \tilde{T}_{i+1}$ is reasonable. The analysis of the error caused by the approximation is shown in the Appendix. Fig.~\ref{fig2} (c) and (d) exhibit that, wave functions extend throughout the whole chain for ergodic states, whereas wave functions are trapped in an interval smaller than $L$ for weakly ergodic states. Note that, wave functions of weakly ergodic states are at the opposite partitions in Fig.~\ref{fig2} (c) and (d), this is because the distribution of weakly ergodic states is energy-dependent.
In the Appendix, we show that the weakly ergodic state exhibits higher than exponential decay in the tail, thus the Lyapunov exponent should be $\gamma(E)\rightarrow\infty$ for the decay distance of the wave function $X_{decay}\rightarrow\infty$. However, due to the fact that $X_{decay}$ is limited by the finite system size, the Lyapunov exponent is finite in the present work. 

\begin{figure}[htbp]
	\includegraphics[width=1.0\columnwidth,height=0.8\columnwidth]{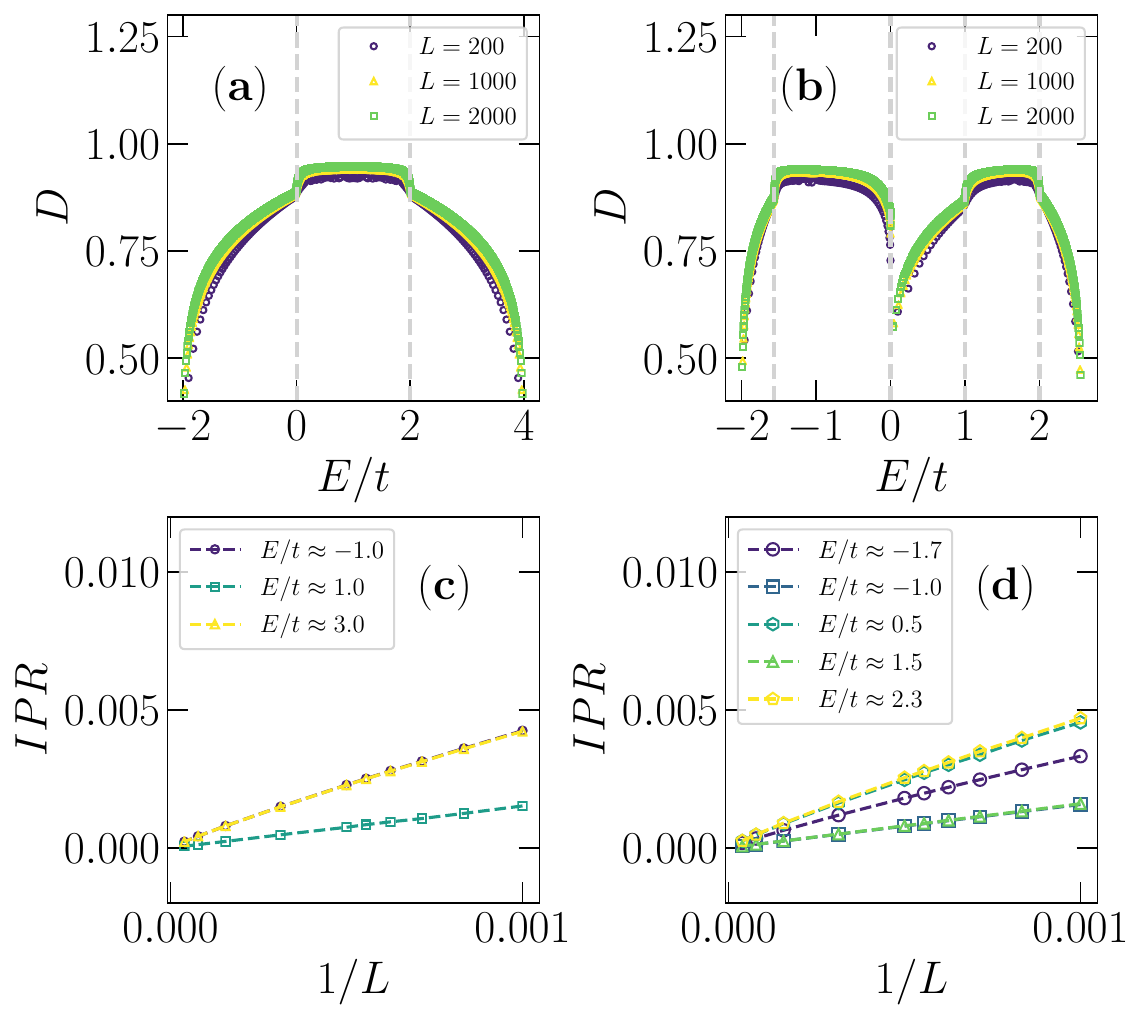}
	\vspace{-0.4cm}
	\caption{ Fractal dimension $D$ and the scaling analysis of IPR for $\kappa=1$  (left panels) and $\kappa=2$  (right panels). The grey dotted lines in (a) and (b) mark the same critical energies as those in Fig.~\ref{fig2} (a) and (b), respectively. The legends in (c) and (d) indicate the target eigenvalue $E$. The largest size in (c) and (d) is $L=25600$. $V_{max}/t=2.0$ in the left panels and $V_{max}/t=1.0$ in the right panels.
    }
	\label{fig3}
\end{figure}

In order to more accurately describe the properties of eigenstates, we further investigate the fractal dimension of the wave function, which is associated with the inverse participation ratio (IPR), $\sum_j|\psi_j|^4$, and defined as
\begin{equation}\label{fd}
\begin{aligned}
D=\frac{-\ln \text{(IPR)}}{\ln L},
\end{aligned}
\end{equation}
In the thermodynamic limit, $D=1$ for ergodic states, $D=0$ for localized states, and $0<D<1$ for fractal states~\cite{EversAnderson,Rosenzweig,Kravtsov,Xianlong2,Ghosh,Khaymovich}, i.e., critical states in other works~\cite{Xin-Chiexact,MirlinExact,JANSSEN,DubertrandTwo}. Evidently, the regions with larger fractal dimensions in Fig.~\ref{fig3} (a) and (b) are consistent with the regions where $\gamma(E)=0$ in Fig.~\ref{fig2} (a) and (b). 
And the discontinuous variations in the derivative of $D$ exactly correspond to the critical energies marked by grey dotted lines. In Fig.~\ref{fig3} (c) and (d), we do the scaling analyses of IPR for ergodic states and weakly ergodic states. By using $IPR\propto(1/L)^D$ to fit the data, one can find that $D=1$ for ergodic states, whereas $D$ is slightly less than one for weakly ergodic states in finite sizes. Referring to Ref.~\cite{AbsenceLonghi,SarmaMobility,Wangone,zhuStatic,Masudul}, one may expect $D=1$ for weakly ergodic states in the thermodynamic limit.  The fractal dimension of weakly ergodic states can also be estimated by assuming the wave function with excitation uniformly distributed over $\Delta$ sites of the lattice~\cite{AbsenceLonghi}. Thus IPR can be estimated by the relation IPR$\sim 1/\Delta$. For $\Delta = fL$ with any finite $0<f<1$, where $f$ is a L-independent prefactor, the fractal dimension can be written as $D=1+\ln(f)/\ln(L)$. In the thermodynamic limit $L\rightarrow\infty$, the fractal dimension $D\rightarrow1$.

\begin{figure}[htbp]
	\includegraphics[width=1.0\columnwidth,height=0.8\columnwidth]{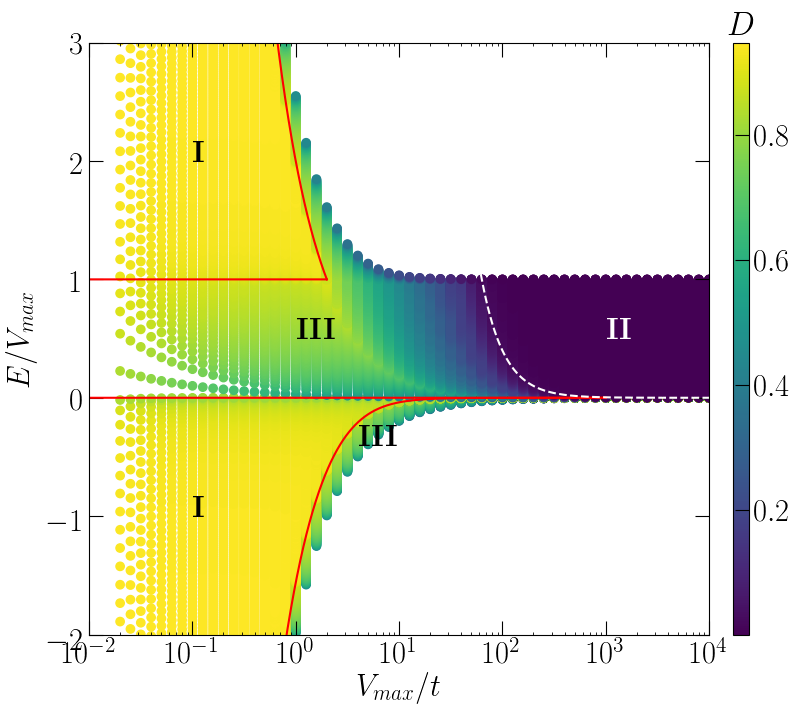}
	\vspace{-0.4cm}
	\caption{The entire phase diagram of the mosaic potential with $\kappa=2$. Different phases are diagnosed by the fractal dimension. Red solid lines and the white dotted line mark critical energies by Eq. \eqref{k2} and Ref.~\cite{Dwiputrasingle}, respectively. Color represents the value of the fractal dimension.  $L=2000$.}
	\label{fig5}
\end{figure}

Consequently, we have determined critical energies separating weakly ergodic states from ergodic states.
Here we complete the entire phase diagram of the mosaic system by referring to the previous work about strongly Wannier-Stark localized states~\cite{Dwiputrasingle}.
It should be clearly pointed out that the Lyapunov exponent cannot distinguish between weakly ergodic and strongly Wannier-Stark localized states, due to $\gamma(E)>0$ for both two states. Although the fractal dimensions of ergodic states and weakly ergodic states are both $D=1$ in the thermodynamic limit, there are differences between them in finite sizes, thus we utilize the fractal dimension to characterize different phases for mosaic potentials.
As shown in Fig.~\ref{fig5}, the red solid lines are the critical energies obtained in the present work,
while the white dashed line is the ``mobility edge'' in Ref.~\cite{Dwiputrasingle}. These two types of critical energies separate the spectra into three regions: $D\approx1$ for the ergodic region (\uppercase\expandafter{\romannumeral1} region), $D\approx0$ for the strongly Wannier-Stark localized region (\uppercase\expandafter{\romannumeral2} region), and the remaining region is the weakly ergodic region (\uppercase\expandafter{\romannumeral3} region).
It is worth mentioning that the recent work shows that, all states are Wannier-Stark localized states with the exception of $(\kappa-1)$-isolated extended states~\cite{AbsenceLonghi}. This conclusion is not contradictory to the present work, since we employ a finite-height Wannier-Stark ladder where $V_{max}$ does not depend on the size, while they use an infinite-height one with $V_{max}\propto L$.
More explicitly, when $L\rightarrow \infty$,  $V_{max}$ is finite in the present work, whereas $V_{max}\rightarrow\infty$ in Ref.~\cite{AbsenceLonghi}. These two types of potentials ($V_{max}$ is finite or infinite) can both be realized in experiments, corresponding to the weak and strong linear potentials~\cite{Mendez,Voisin,Mendez2}, respectively. Furthermore, we use a potential interval as [0, $V_{max}$], whereas Ref.~\cite{AbsenceLonghi} uses a potential interval containing positive and negative values and is symmetric about zero. Crucially, our results show that critical energies highly depend on $V_{max}$ and the system with a finite-height Wannier-Stark ladder is a good platform for the observation of the coexistence of ergodic states and weakly ergodic states.

\begin{figure}[htbp]
	\includegraphics[width=1.0\columnwidth,height=1.2\columnwidth]{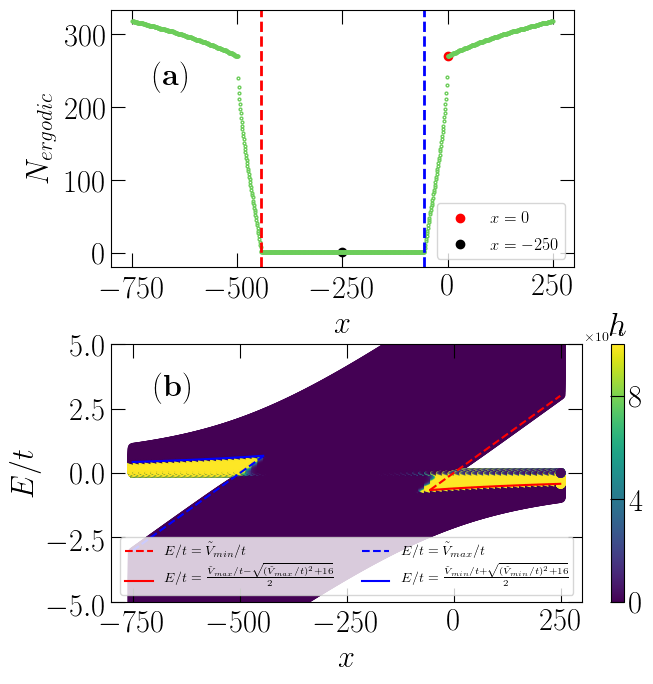}
	\vspace{-0.4cm}
	\caption{(a) The number of ergodic states $N_{ergodic}$ as a function of $x$. (b) The spectrum as a function of $x$. $L=1000$, $\Delta V/t=6$, $\kappa=2$, $V/t=\Delta V/(\kappa(N-1)t)=3/499$. The red and blue dotted lines in (a) indicate $x_{c1}=-444$ and $x_{c2}=-55$, respectively. The black and red solid dots in (a) correspond to the potential settings in Ref.~\cite{AbsenceLonghi} and in Ref.~\cite{Dwiputrasingle}, respectively.
    }
	\label{fig4}
\end{figure}

Above we set the location of the supercell as $i=0, 1, 2, ... , N-1$, here we consider a more general case by setting $i=x, x+1, x+2, ... , x+N-1$. Obviously, altering $x$ does not change the width of the non-zero potential $\Delta V=\kappa V(N-1)$. In Fig.~\ref{fig4}, we fix $\kappa=2$ and $\Delta V/t= 6$ to study the effect of $x$ on the number of ergodic states. By taking $i=x, x+1, x+2, ... , x+N-1$ into the above method of calculating the critical energies, one can obtain that in addition to $E/t=0$, ergodic states can also exist in the energy regions
\begin{equation}
	\left\{\begin{aligned}
		& \tilde{V}_{max}<E<(\tilde{V}_{min}+\sqrt{(\tilde{V}_{min})^2+16})/2, & { x < -\frac{N-1}{2}} \\
		&(\tilde{V}_{max}-\sqrt{(\tilde{V}_{max})^2+16})/2<E<\tilde{V}_{min}, & x \geq -\frac{N-1}{2} \\
	\end{aligned}\right.
\end{equation} 
where $\tilde{V}_{max}=2 V (x+N-1)$ and $\tilde{V}_{min}=2 V x$. By a direct computation,  $\tilde{V}_{max}=(\tilde{V}_{min}+\sqrt{(\tilde{V}_{min})^2+16})/2$ and $(\tilde{V}_{max}-\sqrt{(\tilde{V}_{max})^2+16})/2= \tilde{V}_{min}$ give the critical cases $x_{c1}=4(N-1)/(\Delta V/t)^2-(N-1)$ and $x_{c2}=-4(N-1)/(\Delta V/t)^2$, respectively. Thus, we obtain that ergodic states only survive at $E/t=0$ for $\kappa=2$ for $x_{c1}<x<x_{c2}$. Typically, only one ergodic state survives at $E/t=0$ for $x=-250$. On the contrary, the number of ergodic states is more than one for $x>x_{c2}$ and $x<x_{c1}$. For $x=0$ in Fig.~\ref{fig4}, the number of ergodic states is about 270. To show how the number of the ergodic states changes with $x$ intuitively, we calculate the product of the wave function at the beginning and end of the lattice $h=|\psi_0\psi_{L-1}|$ in  Fig.~\ref{fig4} (b). For ergodic states, one may expect $|\psi_0|>0$ and $|\psi_{L-1}|>0$, resulting in $h>0$. Different from it, $h=0$ for weakly ergodic states and strongly Wannier-Stark localized states. Evidently, different settings of potential intervals have a significant impact on the number of ergodic states. The non-zero potential with zero symmetry is not conducive to the existence of ergodic states.

\section{conclusion}
In summary, we have studied the transition in one-dimensional systems subjected to finite-height linear and mosaic potentials, respectively. In the present work, by exploiting the property that the nearest-neighbor transfer matrices are approximately equal, we introduce a method to obtain the Lyapunov exponent, and exactly determine critical energies separating weakly ergodic states from ergodic states in both cases. Especially, in the latter case, we demonstrate a richer phase diagram relative to the previous work, including an ergodic region, a weakly ergodic region, and a strong Wannier-Stark localization region. We find that critical energies are highly dependent on the height of the ladder and the region of ergodic states is compressed as the height of the ladder increases. Importantly, we find that the number of ergodic states is regulated by the potential interval. By adjusting the potential from symmetry about zero to asymmetry, the number of ergodic states increases significantly. These interesting features will bring new perspectives to a wide range of localization and disorder-free systems.

\begin{acknowledgments}
We acknowledge support from the Natural Science Foundation of China (Grant Nos. 12074340), the Natural Science Foundation of Jiangsu Province (Grant No. BK20200737), the Zhejiang Provincial Natural Science Foundation of China under Grant No.LQ24A040004, and the Science Foundation of Zhejiang Sci-Tech University (Grant Nos. 23062152-Y and 20062098-Y).
\end{acknowledgments}

\bibliography{reference}

\clearpage

\onecolumngrid

\appendix
\begin{appendices}

\vspace{0.3cm}

\twocolumngrid

\beginsupplement

\section{probing by the dynamical evolution }
\begin{figure}[htbp]
	\includegraphics[width=1.0\columnwidth,height=1.2\columnwidth]{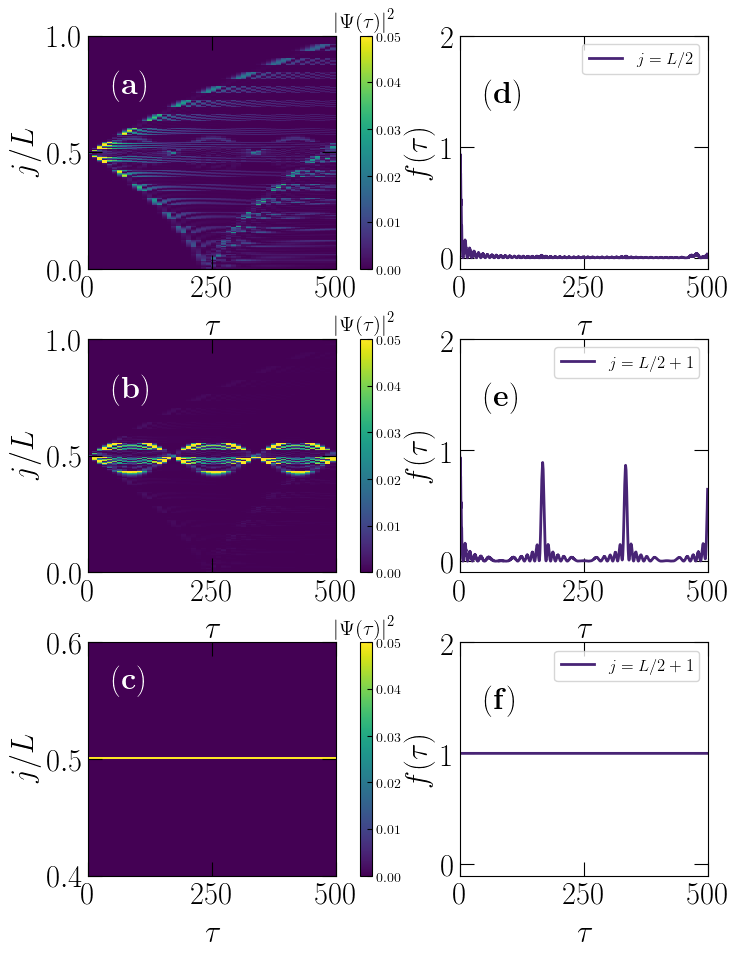}
	\vspace{-0.4cm}
	\caption{The wave functions after time evolution $|\Psi(\tau)|^2$ in the left panels and the corresponding fidelities $f(\tau)$ in the right panels as a function of time $\tau$.
In (a) and (d), the particle is at $j=L/2$ in the initial state, whereas it is at $j=L/2+1$ in (b), (c), (e), and (f).
$V_{max}/t=10$ in (a), (b), (d), and (e), while $V_{max}/t=10^3$ in (c) and (f).
$L=500$, $\kappa=2$.
To indicate the wave functions clearly, we control the range of the color bar from 0.0 to 0.05 in the left panels.
    }
	\label{figadd5}
\end{figure}

To dynamically identify ergodic states, weakly ergodic states, and strongly Wannier-Stark localized states, we investigate the dynamical evolution $|\Psi(\tau)\rangle=e^{-iH\tau}|\Psi(0)\rangle$ of the initial state $|\Psi(0)\rangle$ and the fidelity $f=\left|\langle\Psi(\tau)|\Psi(0)\rangle\right|^ 2$, as done in Ref.~\cite{gaoobservation}.
Firstly, we set $V_{max}=10$ for a weak mosaic potential. In Fig.~\ref{figadd5} (a), the wave function spreads to the entire chain during time evolution when the energy of the initial state is in the ergodic region. On the contrary, Fig.~\ref{figadd5} (b) shows that when the energy of the initial state is in the weakly ergodic region, the wave function oscillates periodically with time. This phenomenon is the well-known  Bloch oscillation~\cite{Bloch}. The fidelities corresponding to the wave functions in Fig.~\ref{figadd5} (a) and (b) manifest that $f(\tau)$ drops to zero after long-time evolution and the local information of the initial state is erased in Fig.~\ref{figadd5} (d), whereas $f(\tau)$ periodically oscillates to preserve the local information in Fig.~\ref{figadd5} (e).
Secondly, we set $V_{max}=10^3$ for a strong mosaic potential to study the dynamical evolution of strongly Wannier-Stark localized states. In Fig.~\ref{figadd5} (c), the wave function does not change obviously during time evolution, because the amplitude of the Bloch oscillation is suppressed. Correspondingly, in Fig.~\ref{figadd5} (f) the fidelity keeps $f\approx1.0$ and the local information is stored for any time.

\section{the boundary effect}
\begin{figure}[htbp]
	\includegraphics[width=1\columnwidth,height=0.8\columnwidth]{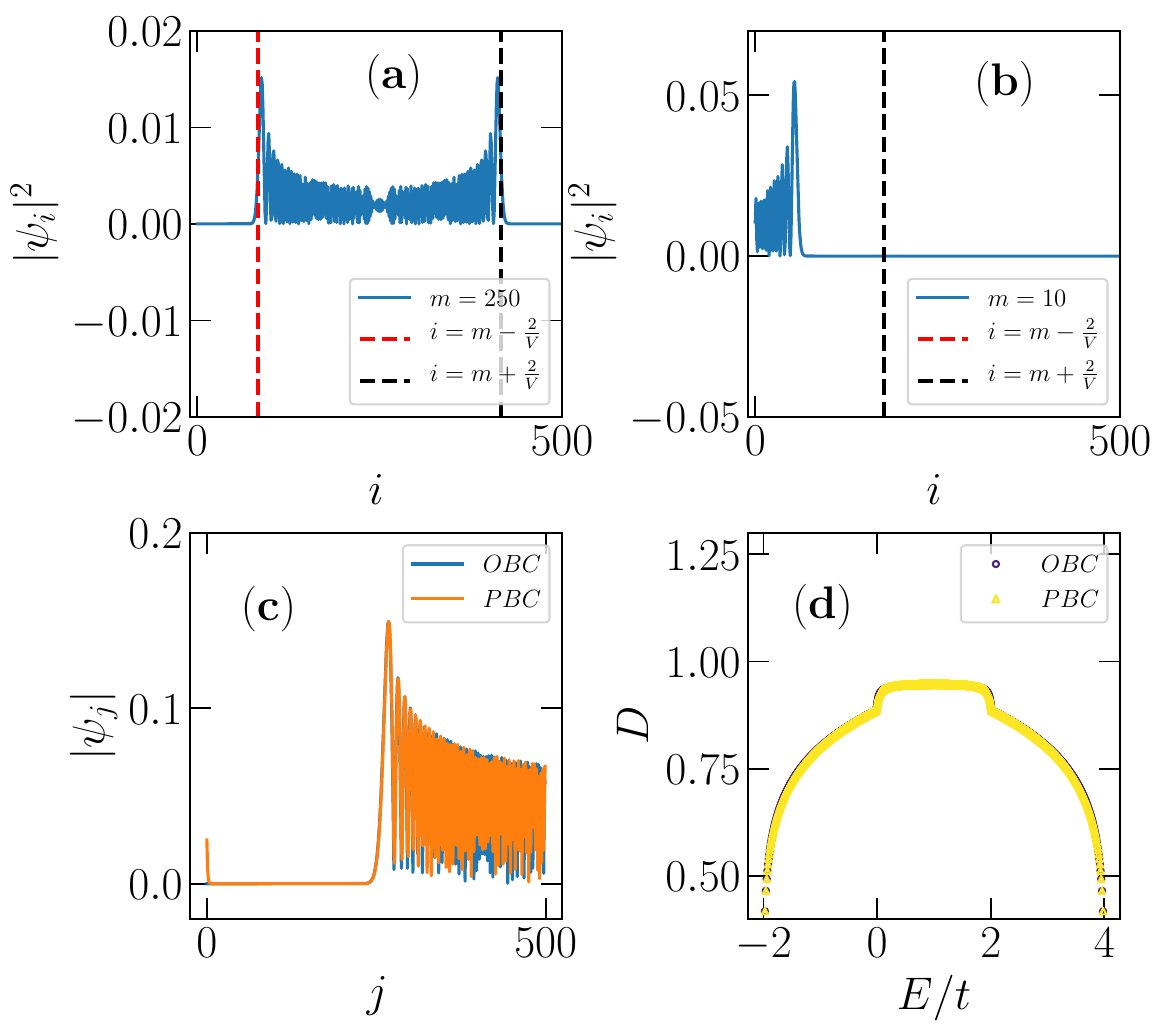}
	\vspace{-0.6cm}
	\caption{(a) and (b) Spatial distribution of the 250th eigenstate and the 10th eigenstate. (c) and (d) Wave functions and fractal dimensions for different boundary conditions. $L=500$, $V_{max}/t=6$ and $\kappa=1$ in (a) and (b). $L=500$. $V_{max}/t=2.0$, $E/t\approx3.048$ in (c), corresponding to Fig.~\ref{fig2} (c) in the main text. $L=2000$, $V_{max}/t=2.0$ in (d), corresponding to Fig.~\ref{fig3} (a) in the main text. For $m=10$, $i=m-\frac{2}{V/t}=m-\frac{2(L-1)}{V_{max}/t}<0$, thus  the red dotted line indicating $i=m-\frac{2}{V/t}$ is not shown in (b).}
	\label{fig_add0}
\end{figure}

To illustrate why we do not use the results in Ref.~\cite{AbsenceLonghi}, we do the following analysis. We start from the transfer matrix in the main text,
\begin{equation}
	\begin{aligned}
		\tilde{T}_i & =\left(\begin{array}{cc}
			E-V\kappa i  & -1 \\
			1 & 0
		\end{array}\right)\left(\begin{array}{cc}
			E & -1 \\
			1 & 0
		\end{array}\right)^{\kappa-1}.
	\end{aligned}
\end{equation}
By using the Sylvester's law, it can be written as
\begin{equation}
	\begin{aligned}
		\tilde{T}_i &=\left(\begin{array}{cc}
			E-V\kappa i & -1 \\
			1 & 0
		\end{array}\right) \left(\begin{array}{cc}
			\frac{\sin (\kappa \omega)}{\sin \omega} & \frac{-\sin [(\kappa-1) \omega]}{\sin \omega} \\
			\frac{\sin [(\kappa-1) \omega]}{\sin \omega} & \frac{-\sin [(\kappa-2) \omega]}{\sin \omega}
		\end{array}\right),
	\end{aligned}
\end{equation}
where the complex angle satisfies $\omega=\arccos\frac{E}{2}$. This transfer matrix can be abbreviated as
\begin{equation}
	\begin{aligned}
	\tilde{T}_i =\left(\begin{array}{ll}
		S_{11} & S_{12} \\
		S_{21} & S_{22}
	\end{array}\right),
	\end{aligned}
\end{equation}
where
\begin{equation}
	\begin{aligned}
		& S_{11}=\frac{\sin (\kappa \omega)}{\sin \omega}\left(E-V\kappa i\right)-\frac{\sin [(\kappa-1) \omega]}{\sin \omega} \\
		& S_{12}=-\frac{\sin [(\kappa-1) \omega]}{\sin \omega}\left(E-V\kappa i\right)+\frac{\sin [(\kappa-2) \omega]}{\sin \omega} \\
		& S_{21}=\frac{\sin (\kappa \omega)}{\sin \omega} \\
		& S_{22}=-\frac{\sin [(\kappa-1) \omega]}{\sin \omega} .
	\end{aligned}
\end{equation}
Utilizing $\operatorname{det} S=S_{11} S_{22}-S_{12} S_{21}=1$, we get the recursive relation of the amplitudes $\psi_i$ as
\begin{equation}
	\begin{aligned}
		\psi_{i+1}+\psi_{i-1}&=\left(S_{11}+S_{22}\right) \psi_i \\
		&=[2 \cos (\kappa \omega)-\frac{\sin (\kappa \omega)}{\sin \omega} V\kappa i]\psi_i.
	\end{aligned}
\end{equation}
For $i\rightarrow\infty$, the above recursive relation  is expressed as
\begin{equation}\label{ss}
	\begin{aligned}
		\psi_{i+1}+\psi_{i-1}&=\frac{\sin (\kappa \omega)}{\sin \omega} V\kappa i\psi_i.
	\end{aligned}
\end{equation}
The recursive relation of Bessel functions is given by
\begin{equation}
	\begin{aligned}
		\mathcal{J}_{i+1}(x)+\mathcal{J}_{i-1}(x)=\frac{2i}{x}\mathcal{J}_{i}(x).
	\end{aligned}
\end{equation}
Taking into account that the wave function is energy level dependent, one can obtain the set of solutions to Eq. \eqref{ss}
\begin{equation}
	\psi_i^{(m)}=(-1)^{i-m} \mathcal{J}_{i-m}(\Gamma),
\end{equation}
where $m$ is the energy level index and
\begin{equation}\label{sss}
	\Gamma=\frac{2}{V\kappa} \frac{\sin \omega}{\sin (\kappa \omega)}.
\end{equation}
The properties of the Bessel functions show that $\mathcal{J}_{i-m}$ is mainly localized in the interval $|i-m|<\Gamma$, which is used to estimate IPR in Ref.~\cite{AbsenceLonghi}. To check the localization properties of the eigenstate, we choose $\kappa=1$ as an example. In this case, $\Gamma$ simplifies to $\Gamma=2/V$, thus the $m$th eigenstate is mainly localized within $(m-2/V)<i<(m+2/V)$, which is verified in Fig.~\ref{fig_add0} (a). However, it can also be found that not all eigenstates satisfy the above localization interval. Typically, we plot the 10th eigenstate for the same parameters in Fig.~\ref{fig_add0} (b), in which the localization interval violates Eq. \eqref{sss} derived from the Ref.~\cite{AbsenceLonghi}. It can be understood as the distribution of the eigenstate is affected by the drop between the end and the head of the lattice chain. Remarkably, the weakly ergodic states, separated from ergodic states by critical energies in our work, are localized at the head or the end of the lattice chain, thus they cannot be described by the analytical solution in Ref.~\cite{AbsenceLonghi} and also do not follow certain conclusions in Ref.~\cite{AbsenceLonghi}. In Fig.~\ref{fig_add0} (c) and (d), we show the wave functions and fractal dimensions for open boundary conditions (OBC) and periodic boundary conditions (PBC), in which one can find that the boundary effect is not obvious.

\section{the scaling behavior of $IPR$ for weakly ergodic states}
\begin{figure}[htbp]
	\includegraphics[width=1\columnwidth,height=0.4\columnwidth]{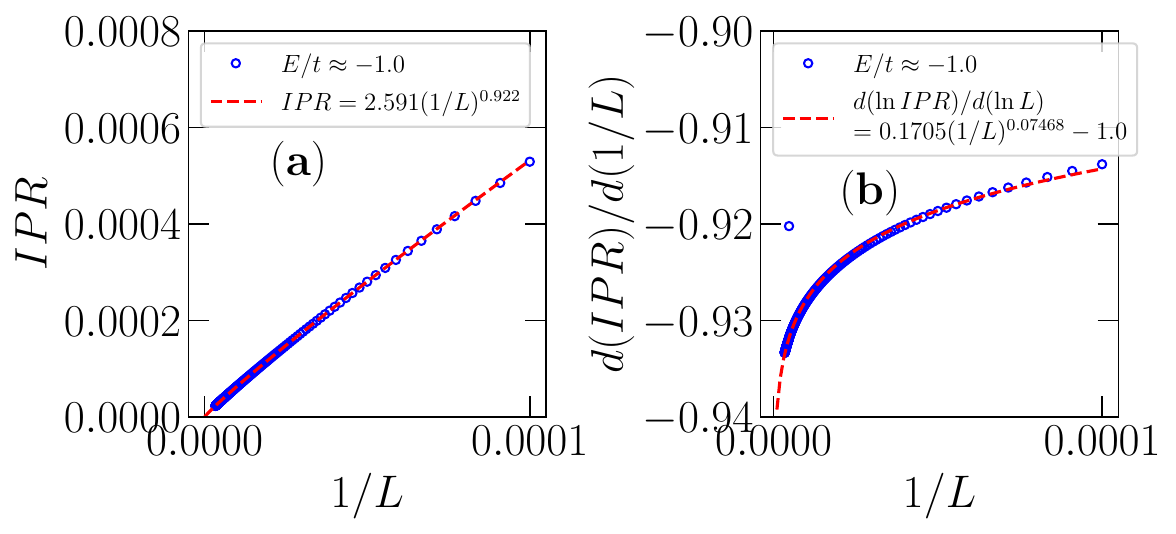}
	\vspace{-0.6cm}
	\caption{(a) $IPR$ versus $1/L$. (b) $d(\ln IPR )/d(\ln L)$ versus $1/L$. $\kappa=1$, $V_{max}/t=2$, the largest size is $L=300000$. Here we use the state-of-the-art shift-invert algorithm.}
	\label{fig_add1}
\end{figure}

In the main text, we have stated that the fractal dimension of weakly ergodic states is slightly less than one for the finite system size. Typically, $D=0.89$ for $E/t\approx-1$ in Fig.~\ref{fig3} (c) in the main text. This fractal dimension is obtained by fitting data with sizes less than $L=25600$. Here we use the state-of-the-art shift-invert algorithm to make the size reach three hundred thousand and fit the data again. Fig.~\ref{fig_add1} (a) indicates $D=0.922$, which is larger than the fractal dimension extracted from small system sizes. This implies that the fractal dimension of the weakly ergodic state requires a very large system size to converge. Intuitively, we show the slope $d(\ln IPR )/d(\ln L)$ as a function of $1/L$ in Fig.~\ref{fig_add1} (b). For ergodic states $IPR\propto(1/L)$, the slope $d(\ln IPR )/d(\ln L)$ should be $-1$, whereas for critical states $IPR\propto(1/L)^D$ with $0<D<1$, the slope $d(\ln IPR )/d(\ln L)$ should be $-D$. In Fig.~\ref{fig_add1} (b), $d(\ln IPR )/d(\ln L)$ decreases as the system size increases, which means that even if the size reaches three hundred thousand, the fractal dimension still does not converge.

\section{the number of different states}
\begin{figure}[htbp]
	\includegraphics[width=1\columnwidth,height=0.8\columnwidth]{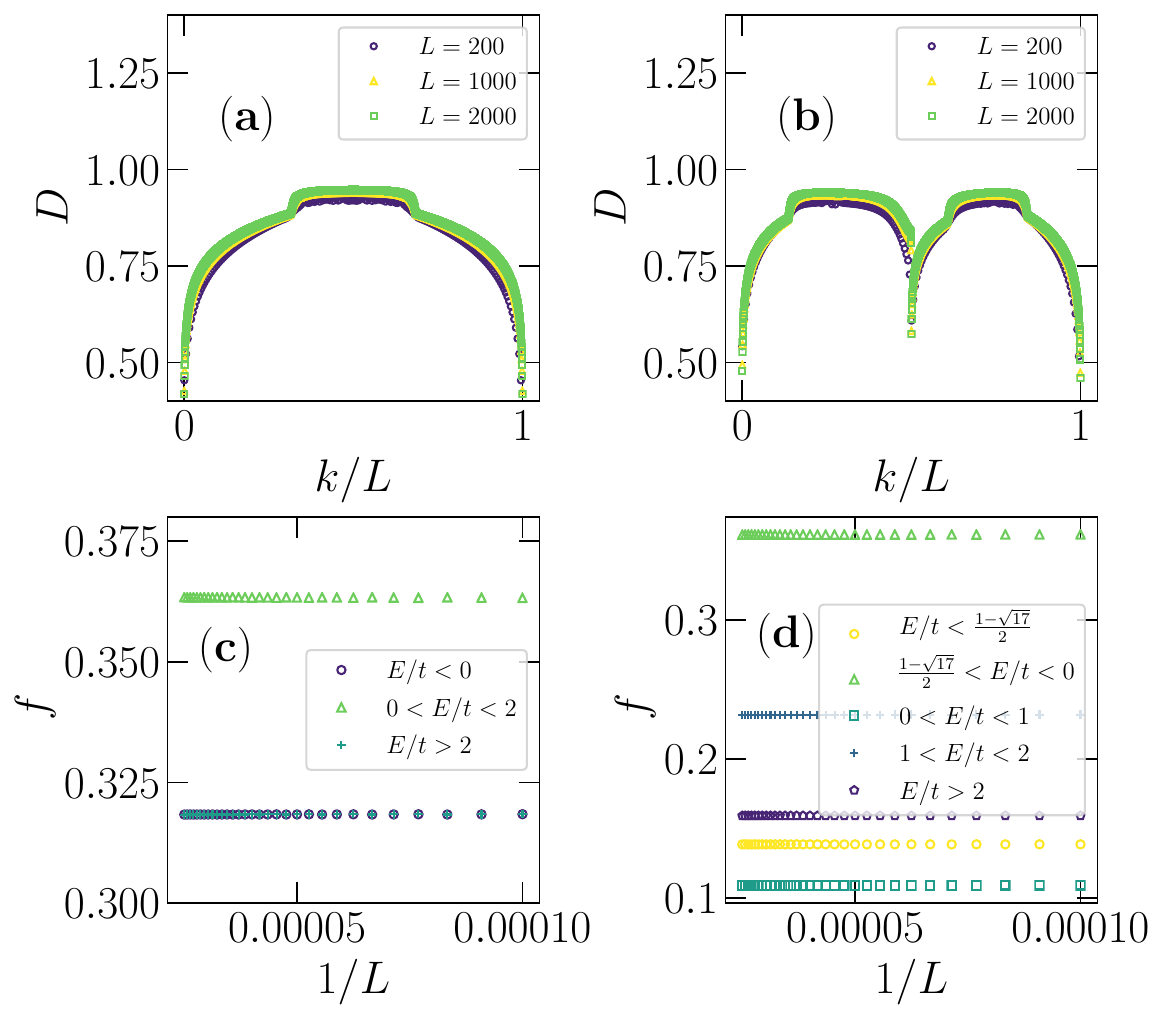}
	\vspace{-0.6cm}
	\caption{Fractal dimension $D$ and the fraction $f$  for $\kappa = 1$ (left panels) and $\kappa=2$ (right panels). In (a) and (b), the abscissa indicates the normalized energy level index, where $k$ is the energy level index. Here, we alter the system size $L$ from $L=10000$ to $L=40000$ in (c) and (d). }
	\label{fig_add2}
\end{figure}

The mobility edge separating extended states and localized states should satisfy that the number of extended states and localized states is a finite fraction of all the states. Strictly speaking, the case in Ref.~\cite{AbsenceLonghi} with only a limited number of extended states and an infinite number of localized states cannot be called mobility edges. Here, we show the fractal dimension $D$ as a function of the normalized energy level index $k/L$ in Fig.~\ref{fig_add2} (a) and (b), where $k$ is the energy level index. These two figures correspond to Fig.~\ref{fig3} (a) and (b) in the main text. One can find that the proportions of ergodic states and weakly ergodic states do not change with size obviously. More intuitively, we show the fraction ($f=N/L$, where $N$ denotes the number of the ergodic (weakly ergodic) states and $L$ is the number of all states) as a function of the system size in Fig.~\ref{fig_add2} (c) and (d). The finite $f$ of weakly ergodic states and ergodic states in the thermodynamic limit indicates that the ratio between them is finite.

\section{the decay of the weakly ergodic states and the site-dependent Lyapunov exponent}
\begin{figure}[htbp]
	\includegraphics[width=1\columnwidth,height=0.8\columnwidth]{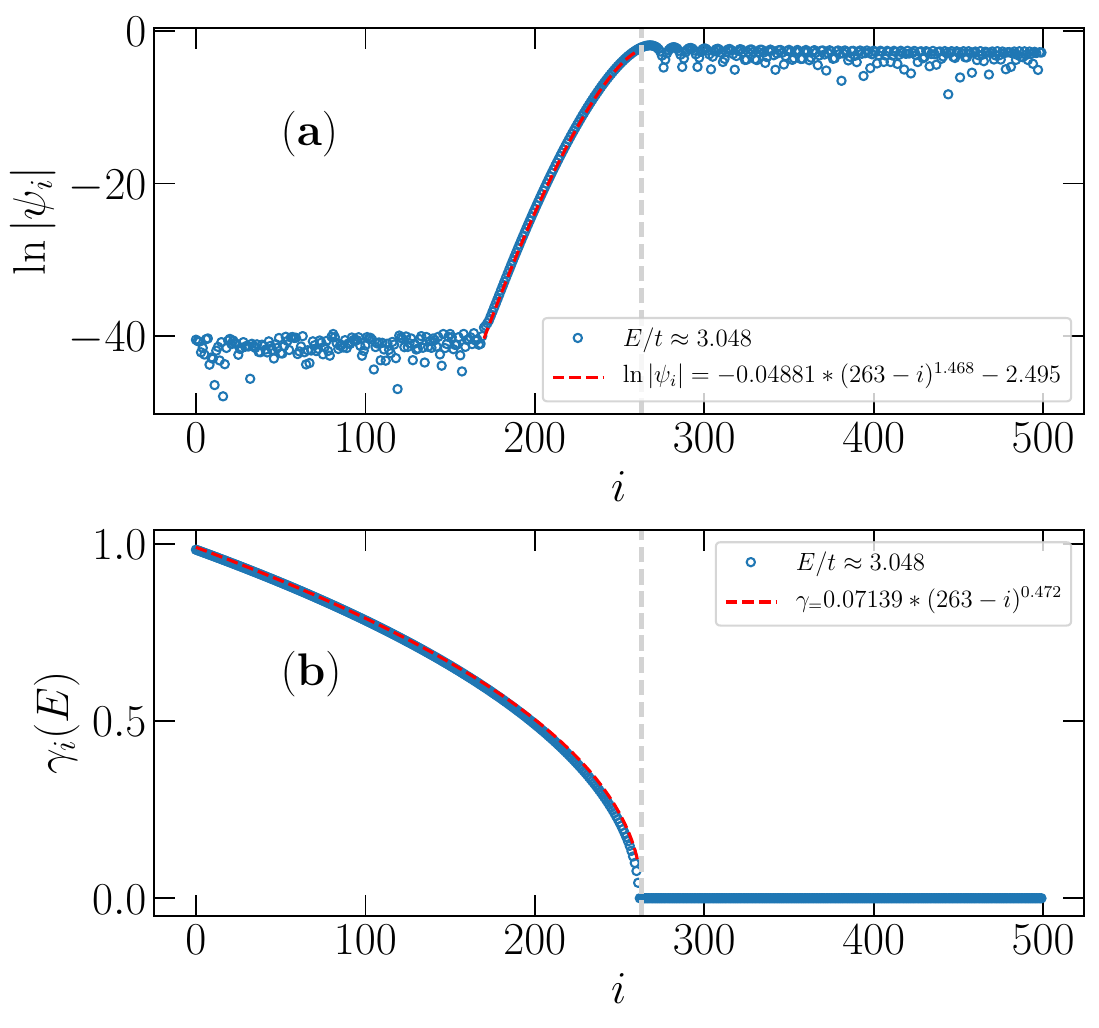}
	\vspace{-0.6cm}
	\caption{(a) $\ln|\psi_i|$ as a function of $i$. (b) The site-dependent Lyapunov exponent $\gamma_i(E)$ as a function of $i$. $\kappa=1$, $L=500$. $V_{max}/t=2.0$. The gray dotted lines in (a) and (b) indicate $i=263$. The value of the wave function for $i<170$ is less than double precision in (a).
}
	\label{fig_add3}
\end{figure}

In Fig.~\ref{fig_add3} (a), we show the typical wave function of weakly ergodic states. The wave function decays as $i$ decreases starting from $i=263$. Here, the value of the wave function for $i<170$ is less than double precision, i.e., $|\psi_i|<10^{-16}$ for $i<170$, thus we choose $170<i<263$ to analyze the wave function. The fitting gives $|\psi_i|=\exp[-0.04881(263-i)^{1.468}-2.495]$, showing a higher than exponential decay. For a wave function decays as $|\psi_x|=\exp[-a x ^b+c]$ ($b>1$), the Lyapunov exponent should be infinity for the decay distance of the wave function $x\rightarrow\infty$. In our work, we find the $\gamma(E)$ is a finite value, this is because $x$ is limited by the finite system size.
In Fig.~\ref{fig_add3} (b), we plot the site-dependent Lyapunov exponent $\gamma_i(E)=\ln||\tilde{T}_i||$ (the Lyapunov exponent of the entire chain can be written as $\gamma(E)=\frac{1}{L}\sum_i\gamma_i(E)$ ). One can find that the non-zero value of $\gamma_i(E)$ is at $i<263$, which is consistent with the region where the wave function decays in Fig.~\ref{fig_add3} (a). For $i>263$, the site-dependent Lyapunov exponent $\gamma_i(E)=0$, agreeing with the region where the wave function extendeds in Fig.~\ref{fig_add3} (a). Here, we also fit $\gamma_i(E)$, which is well-fitted as $\gamma_i(E)=0.07139*(263-i)^{0.472}$.

\section{the error caused by the approximation}
\begin{figure}[htbp]
	\includegraphics[width=1\columnwidth,height=0.4\columnwidth]{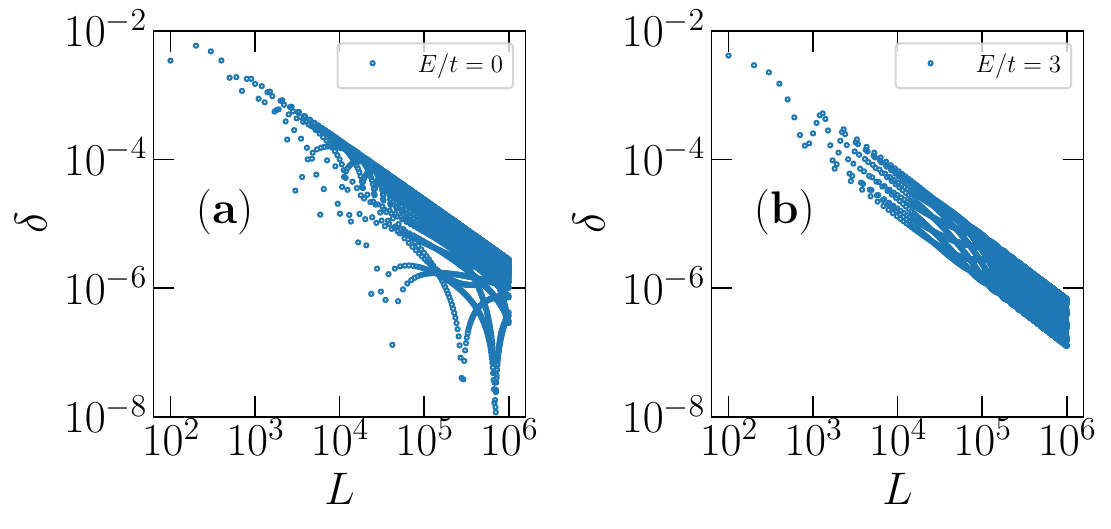}
	\vspace{-0.6cm}
	\caption{The error caused by the approximation for ergodic states and weakly ergodic state. $\kappa=1$, $V_{max}/t=2$. $E/t=0$ in (a) and $E/t=3$ in (b) correspond to the ergodic state and the weakly ergodic state, respectively.}
	\label{fig_add4}
\end{figure}

To determine the error caused by the approximation between Eq. \eqref{e5} and Eq. \eqref{e6}, we define the error as $\delta=\frac{1}{L} \ln (||\prod_{i=0}^{N-1} \tilde{T}_i||)-\frac{1}{L} \ln (\prod_{i=0}^{N-1} ||\tilde{T}_i||)$. In Fig.~\ref{fig_add4} (a) and (b), we show the errors as a function of the system size $L$ for ergodic states and weakly ergodic states, respectively. Although $\delta$ oscillates with the size, one can find that $\delta$ decays exponentially as the size increases, thus it can be expected that $\delta\rightarrow 0$ for $L\rightarrow \infty$. In other words, the approximation is reasonable and Eq. \eqref{e5} and Eq. \eqref{e6} in the main text are equivalent for  $L\rightarrow \infty$.

\end{appendices}
\end{document}